Bartłomiej PŁACZEK[*], Jolanta GOŁOSZ[*]

# THE IN-TOWN MONITORING SYSTEM FOR AMBULANCE DISPATCH CENTRE

The paper presents the vehicles integrated monitoring system giving priorities for emergency vehicles. The described system exploits the data gathered by: geographical positioning systems and geographical information systems. The digital maps and roadside cameras provide the dispatchers with aims for in town ambulances traffic management. The method of vehicles positioning in the city network and algorithms for ambulances recognition by image processing techniques have been discussed in the paper. These priorities are needed for an efficient life-saving actions that require the real-time controlling strategies.

## 1. INTRODUCTION

The integration of informatics, communication and automatic control systems is an indispensable condition for an efficient management and effective activity of rescue services. Fire, traffic accident or fuel outflow are situations, where fast organised rescue actions are expected. Very fast and proper coordinated activities with emergence fast forecast then monitoring these systems, have to be considered.

The camera monitoring system gives a source knowledge about current forces and services available in the moment.

The paper presents several components of monitoring system including the ambulance dispatch centre. The dynamic procedures for emergency vehicles delivering with algorithms for ambulances recognition within urban traffic have been presented.

For localisation of this vehicles the geographical positioning and geographical information systems are recommended.

They are also provided with digital maps and roadside cameras working as traffic recorders. This way the base for dynamic ambulance routing and traffic control strategies are taken under consideration.

---

[*] Silesian University of Technology, Faculty of Transport, Department of Traffic Engineering and Transport Informatics, 40-019 Katowice, ul. Krasińskiego 13. placzek@polsl.katowice.pl

## 2. THE DISPATCH SYSTEM OF IN TOWN AMBULANCE

A quick procedures for finding the nearest free ambulance then the best root selection are only exemplary advantages that the dispatch system offers the modern communication solutions based on GPS and GIS technologies. The dispatch system organisation has been presented by the block scheme in Fig.1

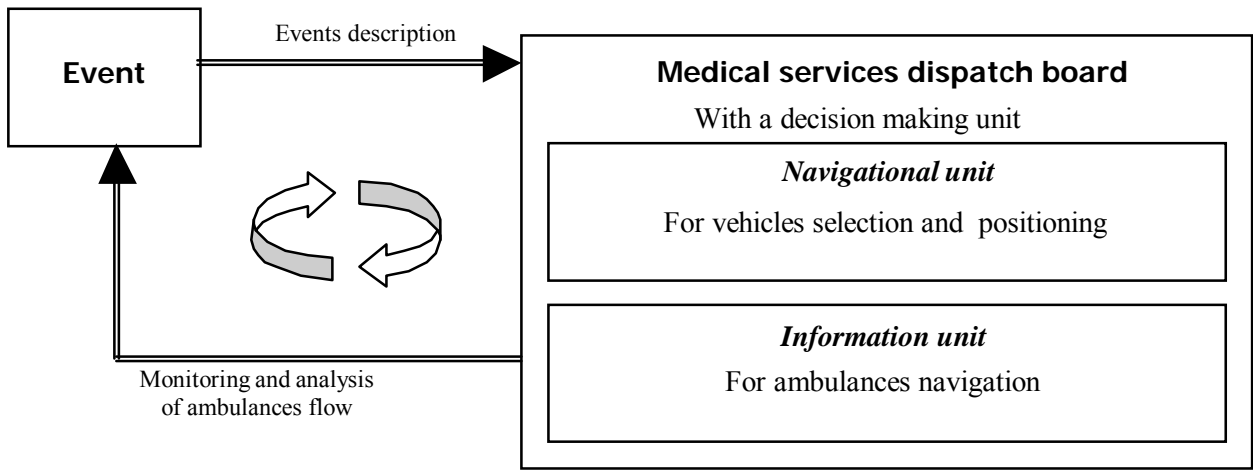

Fig.1. The block scheme of the dispatch system of an ambulances traffic management

The efficient flow of information inside the system is one of fundamental conditions of properly working dispatch board [4]. The dispatch unit analyses a present state of the transportation model then chooses the nearest rescue units sending necessary messages to hospital where the patient can be transported.

The modern dispatch systems are usually supported by video-camera systems used as a traffic recorders, for:
- data collection,
- vehicles priority recognition,
- traffic visualisation,
- finding routes for special vehicles.

This knowledge about traffic state in the network and the ambulance location is a base for decision concerning best root with regard to possible traffic control strategies e.g. the green light route. The major root optimising criterion is time, which is the most important condition of the efficient life-saving actions. The best path can be computed by using Dijkstra's shortest path algorithm [8], where the arcs costs are foresee travel times. The travel times prognosis are determined by an adequate transportation model.

The traffic control system that is implemented for in city roads, has to consider priorities for special vehicles. Main aims of the control strategies is time reduction for travelling. The simplest method of travelling time saving is a green light synchronization on all crossroad lights.

The green light route is defined when the ambulance is recognised in distance about 150 m from crossroad's inlet. Fig. 2a illustrates the state of the ambulance detection during the green signal $G_A$. The duration time $t_d$ of the vehicle access to the stop line is longer then the remaining part of the green signal $G_K$ ($t_d > G_K$). In this case the green light time is elongated by the time $G'_A$, that allows to abandon the crossroad by vehicle /ambulance ($G_K + G'_A > t_d$). The green time can be kept longer over the $G_{max}$, which value is recorded in the traffic local controller.

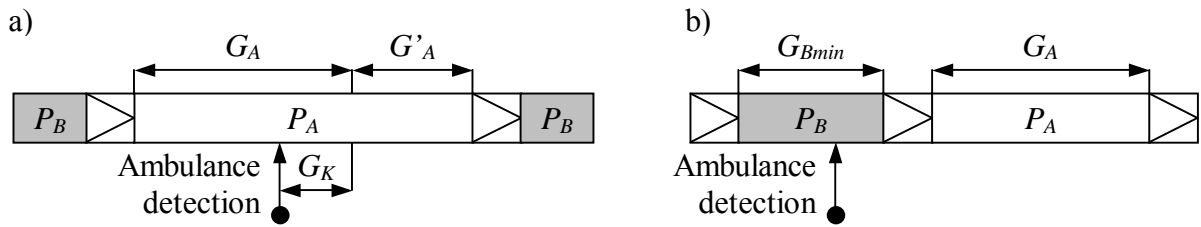

Fig.2. The ambulance priority in traffic signals control

When the ambulance detection takes place during the red light activity (Fig.2b) – in phase $P_B$, the light phases have to be rearranged. Then, the phase $P_B$ is reduced into its minimal value of a green light time $G_{Bmin}$, then the $P_A$ phase is added [3]. It is obvious that this priority for incoming ambulances can not be cancelled by any other vehicles.

The discussed monitoring system can integrate following technologies: GPS, radio communication, GSM, video monitoring and GIS services. The GPS devices deliver the data concerning the vehicles position and their speed (Fig. 3.), as an input of the dispatch functions. The digital map allows to choose an individual route, between two strategic points.

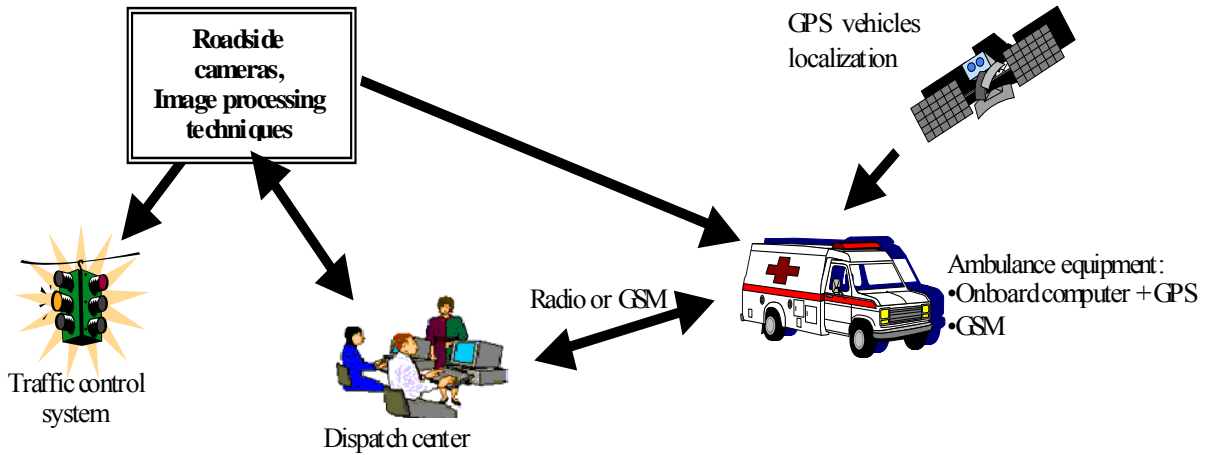

Fig.3. Vehicles monitoring system

## 3. AN ALGORITHM FOR AMBULANCE RECOGNITION BY MEANS OF IMAGE PROCESSING TECHNOLOGIES

The roadside CCD cameras are often used for various vehicles recognition, among them ambulances. Their traffic recognition abilities can be implemented for special vehicles classification. This properties has to be supported by several pre-processing procedures. Fig. 4 shows the block scheme of the ambulance recognition algorithm.

The input data of this algorithm is an image of 8-bit grey scale. An Initial processing of the image obtained from a digital camera consists of two steps: noise reduction using a bend filter then normalization of the image. The normalization process of contrast and brightness provides the intensity function with a whole range of values from 0 to 255. The next step of the image processing uses a Sobel-edge detector that converts the binary image representation into object edges.

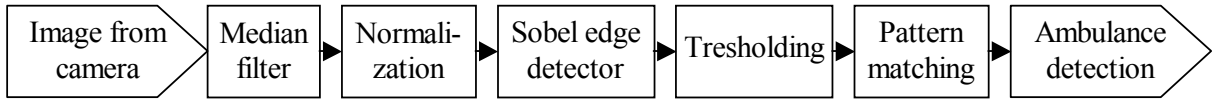

Fig.4. The block scheme of the algorithm for ambulance recognition

In the following operations obtained edge map is searched to find predefined pattern in it. The pattern P and the edge map E are both point sets. Matching the point sets means finding a correspondence between points and minimizing some dissimilarity measure D(P, E) between these sets [6]. Dissimilarity measure is computed for every possible translation of P and the best position, with minimal value D(P, E), is fixed. Thus, this detection method is invariant under translation. In presented algorithm a vehicle from the original image is assumed to be an ambulance if minimal value of dissimilarity measure is small. It means that pattern was find in the edge map.

Let P and E be point sets of sizes n and m respectively. The dissimilarity measure D(P, E) is a sum of distances between points in P and nearest points in E:

$$D(P,E) = \sum_{i=1}^{m} \min_{j=1...m} d(p_i, e_j), \qquad (1)$$

where:
$p_i$, $e_j$ – elements of P and E resp,
$d(p_i, e_j)$ – distance between points $p_i$ and $e_j$.

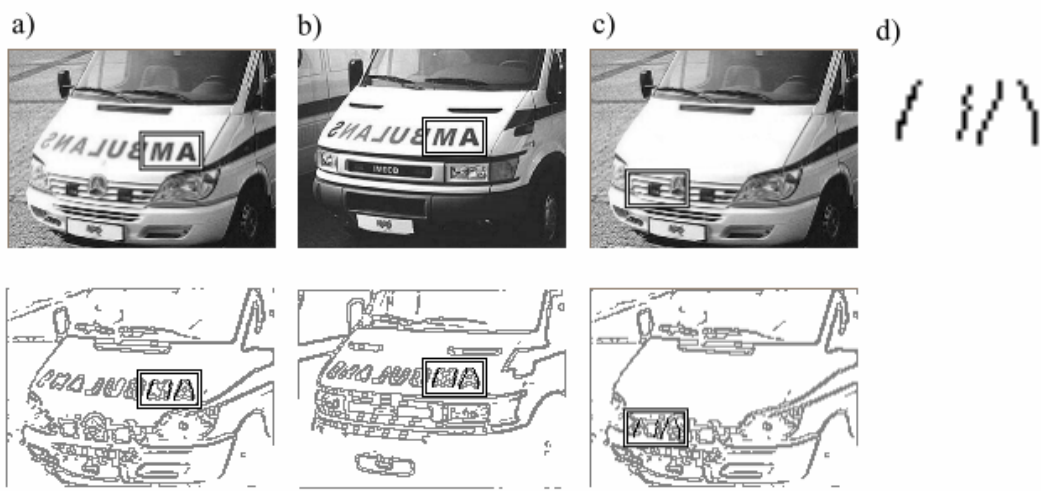

Fig. 5. The example of ambulance identification process
a), b), c) Examined images (upper row) and their edge maps (lower row); d) pattern

For the distance measuring on a discrete grid so-called city block distance definition has been used (formula 2). The city block distance calculation requires lower computational effort then calculation of Euclidean distance and is sufficiently precise for pattern matching purposes. Distance between two points on plane with discrete coordinates $(x_1, y_1)$ and $(x_2, y_2)$ is given by the formula:

$$d((x_1, y_1),(x_2, y_2)) = |x_1 - x_2| + |y_1 - y_2|. \qquad (2)$$

Fig. 5 shows example of ambulance identification process. The experiment was realized for three different images. Images a) and b) present two ambulances, whereas image c) presents the vehicle from picture a) with sign "AMBULANS" removed. The images used in this work are gray scale, 160 by 120 pixels. Edge maps obtained for every image are depicted in lower row in fig. 1. Two first letters of inscription "AMBULANS" on the bonnet was used as identifier of an ambulance. Fig. 5 d) shows pattern that was matched during the detection procedure, its size n is 50 pixels. Rectangular frames in the pictures indicate positions, where the pattern matched best – dissimilarity function $D(P, E)$ had minimal value in these positions.

Various pattern sizes n have been examined (fig. 6). For the pattern size n = 20 pixels the pattern wasn't correctly positioned since number of analyzed pixels was to small in this case. In remaining cases signs "AM" were recognized properly for vehicles a) and b). It is remarkable, that optimal recognition properties of the algorithm were obtained for pattern size n = 50 pixels. For this pattern size it is easy to qualify if a vehicle is an ambulance, basing only on the values of dissimilarity measure.

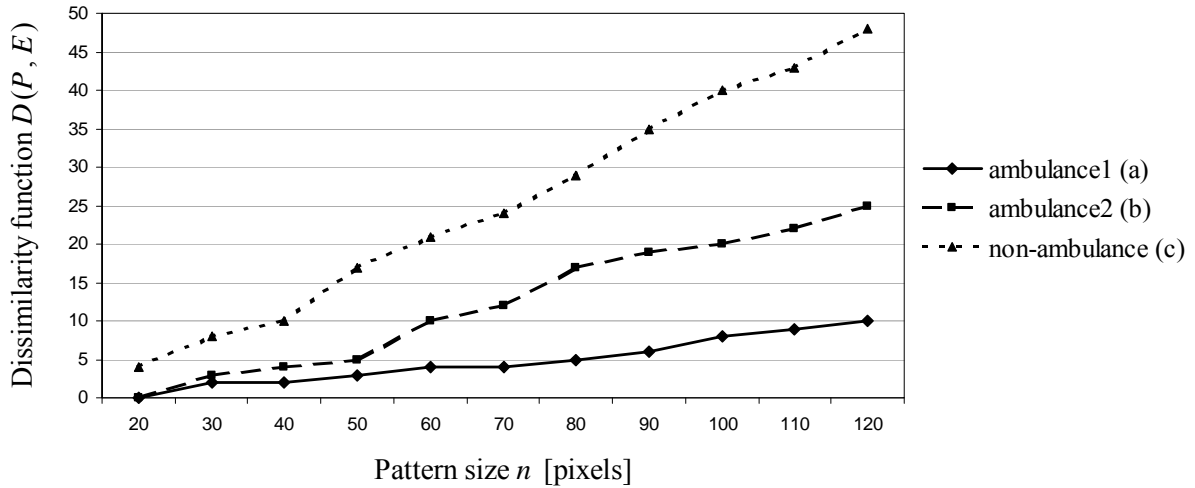

Fig.6. Graph of values of dissimilarity measure for different pattern sizes

The presented examples demonstrate opportunity of an ambulance recognition by image processing techniques. The vehicles stream was divided into two classes: an ambulance or other vehicle, using very simple pattern with four characteristic lines only, recorded by fifty pixels.

Computational complexity of this method is very encouraging what is very important factor for system working in a real-time mode. A proper selection of the pattern is the most important aspect of the defined vehicle extraction. This pattern should contain readable graphics, unique for the given vehicles, not appearing on the other vehicles.

The pattern shape and size determine computational effectiveness in matching all details. More sophisticated pattern models [2] can also be implemented improving the classification reliability.

## 4. CONCLUSIONS

The indicated components of an ambulance recognition and monitoring system can improve the in-town management of emergency vehicles. Utilization of existing systems like GPS and GIS for ambulance dispatch allows to optimize routing with regard to drive time. For the ambulances recognition the roadside cameras are used. The described experiment concerns graphical signs identification placed on a spotted vehicle. They give sufficient input for the vehicle extraction from a traffic stream. The traffic lights control with priority for privileged vehicles are built-in procedures into strategies for emergency services management.